\documentclass[aps,print,showpacs]{revtex4}%
\usepackage{amsfonts}
\usepackage{amsmath}
\usepackage{amssymb}
\usepackage{graphicx}%
\providecommand{\U}[1]{\protect\rule{.1in}{.1in}}

\begin{document}
\title{Nonlinear modes in spatially confined spin-orbit-coupled Bose-Einstein
condensates with repulsive nonlinearity}
\author{Xiong-wei Chen$^{1,\S }$, Zhi-gui Deng$^{2,\S }$, Xiao-xi Xu$^{1,\S }$,
Shu-lan Li$^{1}$, Zhi-wei Fan$^{3}$, Zhao-pin Chen$^{2}$, Bin Liu$^{1}$}
\email{binliu@fosu.edu.cn}
\author{Yong-yao Li$^{1,2}$}
\affiliation{$^{1}$School of Physics and Optoelectronic Engineering, Foshan University,
Foshan 528000, China}
\affiliation{$^{2}$ Department of Physical Electronics, School of Electrical Engineering,
Faculty of Engineering, Tel Aviv University, Tel Aviv 69978, Israel}
\affiliation{$^{3}$ Department of Physics, University of Bath, Bath BA2 7AY, UK}
\affiliation{$^{\S }$ These authors contributed equally to this work.}

\begin{abstract}
It was found that spatially confined spin-orbit (SO) coupling, which can be
induced by illuminating Bose-Einstein condensates (BECs) with a Gaussian laser
beam, can help trap a spinor Bose gas in multi-dimensional space. Previous
works on this topic were all based on a Boson gas featuring an attractive
interaction. In this paper, we consider the trapping effect in the case in
which the Boson gas features a repulsive interaction. After replacing the
repulsive effect, stable excited modes of semi-vortex (SV) type and mixed-mode
(MM) type, which cannot be created in a boson gas with attractive
interactions, can be found in the current setting. The trapping ability and
the capacity of the confined SO coupling versus the degree of the repulsive
strength as well as the order of the excited mode are systematically discussed
firstly through the paper. Moreover, the stability of the nonlinear mode
trapped in this system with a moving reference frame is also discussed. Unlike
the system with homogeneous SO coupling, two different types of stationary
mobility modes can be stabilized when the SO coupling moves in the $x$- and
$y$- directions, respectively. This finding indicates that the system with
moving confined SO coupling features a typical anisotropic character that
differs from the system with moving homogeneous SO coupling.

\end{abstract}
\maketitle

\section{Introduction}

Spin-orbit (SO)-coupled Bose-Einstein condensates (BECs) provide an ideal and
clean platform to emulate the relativistic dynamics of electrons in condensate
matter physics
\cite{Stanescu2008,YJLin2011,ZwuS,Deng2012,Huang2016,Yongping2012,YunLi2013,WBPRA87,RFZ1,RFZ2,WLActa
Physica Sinica 68,L. WenPRA86,WLAnnals of
Physics390,ZXFPRA95,ZXFPRL121,Physics Letters A 383,WJG,PRA97_063607,lbNJP}.
Many novel phenomena, such as topological insulators, superconductors,
supersolids, and the spin-Hall effect \cite{MZH,xlq,JLi2017,Stuhl2015}, can be
simulated by BECs with SO coupling \cite{Zhaihui,Yongping}. Because a boson
gas can feature abundant interactions, it provides many new avenues for people
to reconsider these problems. Moreover, it was also reported that the SO
coupling effect can help an attractive boson gas form stable matter-wave
solitons in two- and three-dimensional (2D and 3D) free space
\cite{Lobanov2014,Sakaguchi2014,Sakaguchi2016,Sakaguchi2018,Malomed2018,Guihua2017,Gautam2017,Sakaguchi2019,3Dsoc}%
. In the area of nonlinear physics, it is well known that free-space solitons
may collapse in 2D and 3D geometries via the action of the usual attractive
cubic nonlinearity \cite{Malomed2005,Berge1998}. Hence, how to stabilize
solitons in multi-dimensional space remains a challenging issue. Generally, a
common operation to stabilize solitons in free space beyond 1D is to modify
the attractive cubic nonlinearity, which includes reducing the cubic type to
the quadratic type \cite{Torruellas,Xliu2000}, changing it to the saturable
nonlinearity \cite{Segev1994}, introducing a nonlocal nonlinearity
\cite{Peccianti2002,Pedri2005,Tikonenkow2008,Jiasheng,Maucher2011}, and adding
competitive nonlinearities with orders different from cubic (such as the
competing cubic-quintic nonlinearity
\cite{Mihalache2001,CQ2,ELF2013,WYYDN90,DCQND87,ND93_2379,ND91_757}).
Recently, a new type of self-bound quantum liquid, named quantum droplets, has
been created from 1D to 3D space via binary BECs
\cite{Petrov2015,Petrov2016,Cabrera2018,BinLiu,YVK2019,Zouzheng,GEA2018} and
dipolar BECs \cite{Schmitt2016,Chomaz,Baillie2016,Wachtler,Ferrier2016,Edler}
with the help of Lee-Huang-Yang (LHY) correction \cite{LHY}. Quantum droplets
with novel vortices in 2D space are also predicted for binary BECs assisted by
the LHY term \cite{Yongyao2018,Yongyao2019,Tengstrand2019}. However, SO
coupling stabilizes solitons in a different way, which are modified via a
linear effect rather than a nonlinear effect. Therefore, SO coupling provides
a new way for people to study solitons in multi-dimensional space.

With the help of SO coupling, families of composite matter-wave solitons, of
the semi-vortex (SV) type and mixed-mode (MM) type, are created in 2D and 3D
space if the spinor BECs feature an attractive interaction. Moreover,
anisotropic stripe solitons and vortex solitons are created with the SV form
in SO-coupled dipolar BECs
\cite{Xuyong2015,Xunda2016,Yongyao20172,Bingjin2017,Bingjin2018,Shimei2018,Weipang2018}%
. Self-trapped modes formed by the interplay of SO coupling and the LHY term
have also been considered \cite{Cuixiaoling,Yongyao2017,Sahu2019}. In optics,
spatiotemporal solitons with an SO coupling-like effect were created in planar
dual-core waveguides and twisted cylinder waveguides with the self-focusing
Kerr nonlinearity \cite{YVK20151,YVK20152,Sakaguchi20162,Haohuang}. However,
these solitons are all ground-state solutions of the system. Even though the
excited-state solitons of these two types of soliton were also predicted, they
are all unstable in the background of the attractive cubic nonlinearity
\cite{Sakaguchi2018}. Recent studies reveal that excited-state solitons can be
stabilized if the attractive nonlinearity is replaced by the repulsive one
\cite{SakaguchiFOP}. However, such a repulsive nonlinearity cannot be
homogeneous; otherwise, the self-trapped mode cannot be formed. References
\cite{Chunqing2018,Rongxuan2018} have reported that stable excited-state
solitons can be found when the intensity of the local or nonlocal repulsive
nonlinearity grows (with a strong growth rate) from the center. Such specific
designs may bring about great difficulty in terms of experimental achievement.
Hence, a natural consideration is whether the excited-state solutions can be
realized under a homogeneous repulsive nonlinear background.

Very recently, it was reported that matter-wave solitons can be trapped if SO
coupling, obtained under illumination by an external laser field, is applied
in a spatially confined area
\cite{YVK2014,Yongyao20192,Zhijiang,DCQND88(2629)}. This finding implies that
if the strength of the SO coupling was spatially modified to be a confined
one, it would be possible to achieve stable excited state modes formed via the
homogeneous repulsive nonlinearity. Based on such speculation, the first
objective of this paper is to demonstrate this possibility. The second
objective of this paper is to verify the trapping capacity of such type of SO
coupling if this possibility has been demonstrated. Then, based on the above
two objectives, the final objective of this work is to consider the dynamics
of these trapped modes if the SO coupling is put into a moving reference
frame. Note that the stability of nonlinear modes with repulsive nonlinearity
in a moving reference frame in 2D geometries has not yet been discussed.
Moreover, this discussion is nontrivial because SO coupling violates Galilean
invariance. To realize these targets, the rest of the paper is structured as
follows. The model is introduced in Section II. Basic numerical results for
the nonlinear mode in quiescent and moving reference frames are reported in
Section III and IV. The paper is concluded in Section V.

\begin{figure}[ptb]
{\includegraphics[width=0.3\columnwidth]{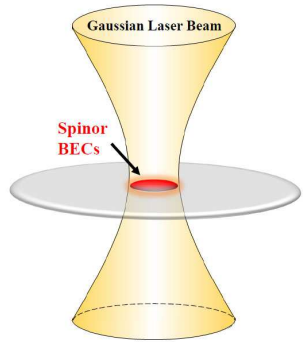}}\caption{(Color online)
Sketch map of the model. Spinor BECs are trapped in a 2D space and illuminated
by a Gaussian laser beam, which creates a spatially confined SO coupling
effect.}%
\label{sketch}%
\end{figure}

\section{The Model}

Similar to the settings in Ref. \cite{Yongyao20192}, a spinor BEC is trapped
by 2D SO coupling of the Rashba type within a confined area, which is
illuminated by a Gaussian laser beam (see Fig.\ref{sketch}). The illuminated
area is also defined by a Gaussian function as
\[
\lambda(r)=\lambda_{0}\exp\left[  -\left(  {\frac{r}{D}}\right)  ^{2}\right]
,
\]
where $\lambda_{0}$ is the strength of the SO coupling and $D$ stands for the
size of the confined area. The mean-field model of this system is based on the
Lagrangian%
\begin{align}
\mathcal{L} &  =-\frac{i}{2}\left(  \Psi_{+}^{\ast}\frac{\partial\Psi_{+}%
}{\partial t}+\Psi_{-}^{\ast}\frac{\partial\Psi_{-}}{\partial t}-c.c.\right)
\nonumber\\
&  +\frac{1}{2}\left(  {{{\left\vert \nabla{{\Psi_{+}}}\right\vert }^{2}%
}+{{\left\vert {\nabla{\Psi_{-}}}\right\vert }^{2}}}\right)  +g\left[
\frac{1}{2}\left(  {{{{\left\vert {{\Psi_{+}}}\right\vert }^{4}}+{{\left\vert
{{\Psi_{-}}}\right\vert }^{4}}}}\right)  {+\gamma{{\left\vert {{\Psi_{+}}%
}\right\vert }^{2}}{{\left\vert {{\Psi_{-}}}\right\vert }^{2}}}\right]
\nonumber\\
&  +\frac{{\lambda(r)}}{2}\left\{  {\left[  {\Psi_{+}^{\ast}\frac
{{\partial{\Psi_{-}}}}{{\partial x}}-\Psi_{-}^{\ast}\frac{{\partial{\Psi_{+}}%
}}{{\partial x}}-i\left(  {\Psi_{+}^{\ast}\frac{{\partial{\Psi_{-}}}%
}{{\partial y}}+\Psi_{-}^{\ast}\frac{{\partial{\Psi_{+}}}}{{\partial y}}%
}\right)  }\right]  +c.c.}\right\}  ,\label{L_density}%
\end{align}
and
\begin{equation}
L=\int{\int}\mathcal{L}dxdy,\label{Lagrangian}%
\end{equation}
where $\Psi_{\pm}$ are the pseudo-spinor wave functions of the condensates,
$c.c.$ and the asterisk in $\mathcal{L}$ stand for the complex conjugate, and
$g$ and $\gamma$ are the total nonlinear strength and the relative strength of
the cross-interaction, respectively. Here, the strength of the
self-interaction is scaled by $1$. In Ref. \cite{Yongyao20192}, $g$ and
$\gamma$ are tuned to feature an attractive interaction. However, in the
current setting, in order to create a stable excited mode, we replace both of
them by a repulsive interaction. It's well known that the scattering length of
the BEC atoms can be either positive (corresponding to a repulsive effective
interaction) or negative (corresponding to an attractive one), which can be
tuned by the Feshbach resonance
\cite{Chin2010,FB2006,Inuye1998,Timmermans1999}.{\LARGE  }The Gross-Pitaevskii
equation (GPE) with the mean-field approximation can be derived from
Lagrangian Eq. (\ref{Lagrangian}) by using the Euler-Lagrange equations as
follows:%
\begin{align}
i\frac{\partial{\Psi}_{+}}{\partial t} &  =-\frac{1}{2}\nabla^{2}{\Psi}%
_{+}+g\left(  \left\vert \Psi_{+}\right\vert ^{2}+\gamma\left\vert \Psi
_{-}\right\vert ^{2}\right)  \Psi_{+}+\lambda(r)\left(  \frac{\partial
}{\partial x}-i\frac{\partial}{\partial y}\right)  \Psi_{-}+\frac{1}%
{2}e^{-i\theta}{\frac{d\lambda}{dr}}\Psi_{-},\nonumber\\
i\frac{\partial{\Psi}_{-}}{\partial t} &  =-\frac{1}{2}\nabla^{2}{\Psi}%
_{-}+g\left(  \left\vert \Psi_{-}\right\vert ^{2}+\gamma\left\vert \Psi
_{+}\right\vert ^{2}\right)  \Psi_{-}-\lambda(r)\left(  \frac{\partial
}{\partial x}+i\frac{\partial}{\partial y}\right)  \Psi_{+}-\frac{1}%
{2}e^{i\theta}{\frac{d\lambda}{dr}}\Psi_{+}.\label{Model}%
\end{align}

Stationary solutions to Eqs. (\ref{Model}) with chemical potential $\mu$ are
sought as follows:
\begin{equation}
{{\Psi_{\pm}}\left(  {\mathbf{r},t}\right)  ={\phi_{\pm}}\left(  {\mathbf{r}%
}\right)  {e^{-i\mu t},}} \label{ansatz}%
\end{equation}
where $\mathbf{r}=x\hat{\mathbf{i}}+y\hat{\mathbf{j}}$ and the functions
${{\phi_{\pm}}\left(  {\mathbf{r}}\right)  }$ satisfy the equations%
\begin{align}
&  \mu\phi_{+}=-\frac{1}{2}\nabla^{2}{\phi}_{+}+g\left(  \left\vert \phi
_{+}\right\vert ^{2}+\gamma\left\vert \phi_{-}\right\vert ^{2}\right)  {\phi
}_{+}+\lambda(r)\left(  \frac{\partial}{\partial x}-i\frac{\partial}{\partial
y}\right)  \phi_{-}+\frac{1}{2}e^{-i\theta}{\frac{d\lambda}{dr}}\phi
_{-},\nonumber\\
&  \mu\phi_{-}=-\frac{1}{2}\nabla^{2}{\phi}_{-}+g\left(  \left\vert {\phi}%
_{-}\right\vert ^{2}+\gamma\left\vert {\phi}_{+}\right\vert ^{2}\right)
{\phi}_{-}-\lambda(r)\left(  \frac{\partial}{\partial x}+i\frac{\partial
}{\partial y}\right)  {\phi}_{-}-\frac{1}{2}e^{i\theta}{\frac{d\lambda}{dr}%
}{\phi}_{-}. \label{Model_phi}%
\end{align}

The energy corresponding to Lagrangian (\ref{L_density}) is
\begin{align*}
E  &  =\int{\int}\left(  \varepsilon_{\text{K}}+\varepsilon_{\text{N}%
}+\varepsilon_{\text{SOC}}\right)  dxdy,\\
\varepsilon_{\text{K}}  &  =\frac{1}{2}\left(  {{{\left\vert \nabla{{\phi_{+}%
}}\right\vert }^{2}}+{{\left\vert {\nabla{\phi_{-}}}\right\vert }^{2}}%
}\right)  ,\\
\varepsilon_{\text{N}}  &  =\frac{g}{2}\left[  \left(  {{{{\left\vert
{{\Psi_{+}}}\right\vert }^{4}}+{{\left\vert {{\Psi_{-}}}\right\vert }^{4}}}%
}\right)  {+2\gamma{{\left\vert {{\Psi_{+}}}\right\vert }^{2}}{{\left\vert
{{\Psi_{-}}}\right\vert }^{2}}}\right]  ,\\
\varepsilon_{\text{SOC}}  &  =\frac{{\lambda(r)}}{2}\left\{  {\left[
{\Psi_{+}^{\ast}\frac{{\partial{\Psi_{-}}}}{{\partial x}}-\Psi_{-}^{\ast}%
\frac{{\partial{\Psi_{+}}}}{{\partial x}}-i\left(  {\Psi_{+}^{\ast}%
\frac{{\partial{\Psi_{-}}}}{{\partial y}}+\Psi_{-}^{\ast}\frac{{\partial
{\Psi_{+}}}}{{\partial y}}}\right)  }\right]  +c.c.}\right\}  ,
\end{align*}
where $\varepsilon_{\text{K}}$, $\varepsilon_{\text{N}}$, and $\varepsilon
_{\text{SOC}}$ are the kinetic, interaction, and SO coupling energy densities, respectively.

In the following study, we will apply the normalized condition and define the
total norm as%
\begin{equation}
N=\int{\int}n\left(  \mathbf{r}\right)  dxdy\equiv1, \label{N}%
\end{equation}
where $n\left(  \mathbf{r}\right)  ={{{{\left\vert {{\phi_{+}}}\right\vert
}^{2}}+{{\left\vert {{\phi_{-}}}\right\vert }^{2}}}}$ is the effective density
distribution of the nonlinear mode. Estimation of the actual number of atoms
for the normalization condition on the real experiment scale can be done: if
the effective length of the SO coupling is $\sim0.5$ m, which corresponds to a
physical length of approximately $1$ $\mu$m, and $N=1$ corresponds to a number
of atoms of $\sim10^{4}$ \cite{Yongyao20192,Zhijiang}.

The stationary solutions of the nonlinear mode of Eqs. (\ref{Model_phi}) are
solved numerically by means of the imaginary-time-integration method (ITM)
\cite{ITP1,ITP2}, and the stabilities of these nonlinear modes are verified
via direct simulations. In the numerical simulations, we normalized
$\lambda_{0}=1$ and leave $(D,g,\gamma)$ as a set of controlled parameters of
the system.

\section{Nonlinear mode in the quiescent reference frame}

\begin{figure}[tbh]
{\includegraphics[width=16cm]{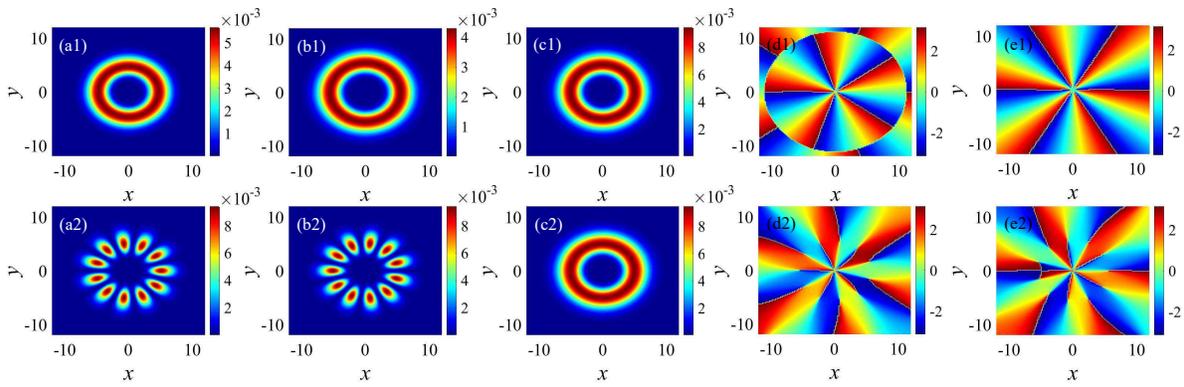}}\caption{(color online) (a1, b1) and
(a2, b2): the density patterns of the $\phi_{+}$ and $\phi_{-}$ components of
the excited SV and MM modes with $S_{+}=5$ and $S_{1}=5$, respectively. (c1,
c2): the total density patterns of the binary BEC. The fourth and fifth
columns: the corresponding phase patterns of $\phi_{+}$ and $\phi_{-}$. Here,
we fixed parameters $D=10$, $g=2$ and $\gamma=1$.}%
\label{ESVEMM}%
\end{figure}

\begin{figure}[tbh]
{\includegraphics[width=15cm]{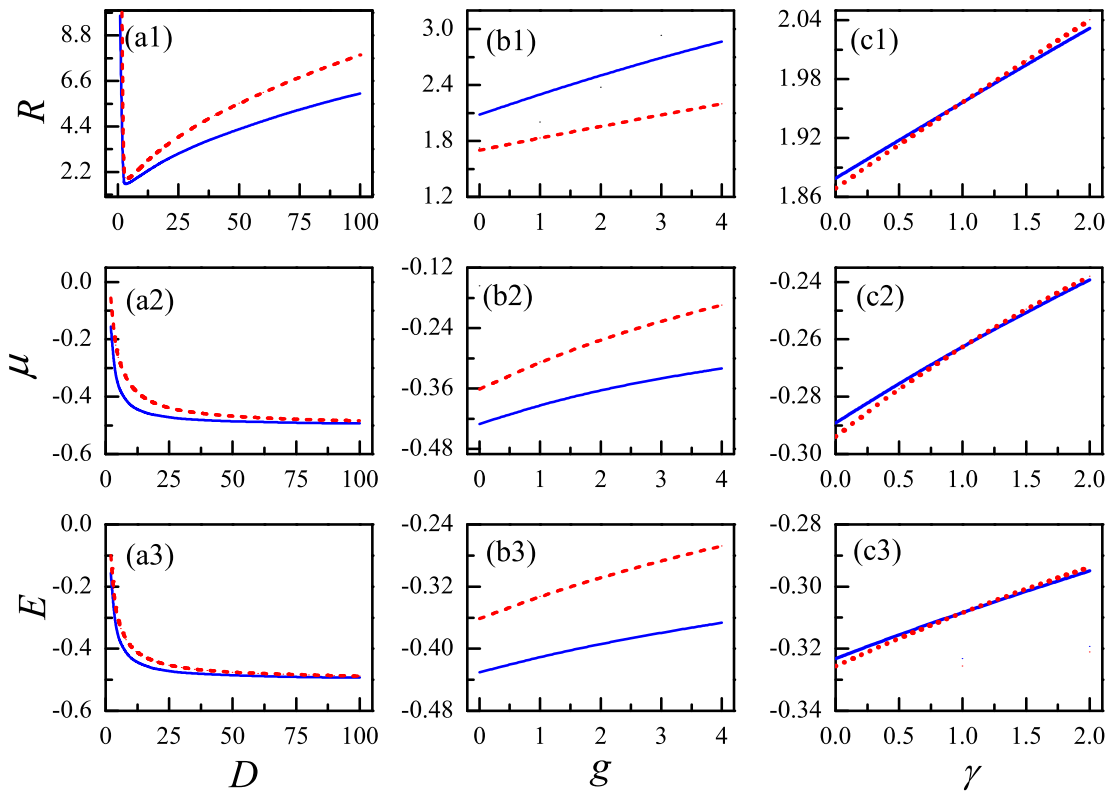}}\caption{(color online) The first
column: the size $R$, the chemical potential $\mu$,\ and the total energy $E$
of the fundamental SVs as functions of the SO coupling confinement size, $D$,
for $g=0$ (the blue line) and $g=2$ (the red short dashed curves). The second
column: the $R$, $\mu$,\ and $E$ of the fundamental SVs as functions of the
total nonlinear strength, $g$, for $D=5$ (the red short dashed curves) and
$D=10$ (the blue line). Here, we fixed $\left(  N,\gamma\right)  =\left(
1,1\right)  $. The last column: the $R$, $\mu$,\ and $E$ of the SVs (the blue
line) and MMs (the red short dashed curves) as functions of the strength of
$\gamma$. Here, we fixed $\left(  N,g,D\right)  =\left(  1,2,10\right)  $. }%
\label{char3}%
\end{figure}

We note that Eqs. (\ref{Model}) admits two types of 2D stationary solutions in
the form of semi-vortex (SV) modes and mixed modes (MMs). According to Refs.
\cite{Chunqing2018,Rongxuan2018}, the fundamental SV modes and their excited
states can be produced through the ITM by inputting the following ansatz:
\begin{equation}
\phi_{\pm}^{(0)}=A_{\pm}r^{S_{\pm}}\exp\left(  -\alpha_{\pm}r^{2}+iS_{\pm
}\theta\right)  ,\label{ESSV}%
\end{equation}
where $A_{\pm}$ and $\alpha_{\pm}$ are positive real constants and $S_{\pm}$
is the vorticity topological charge exerted on the input. Fundamental SVs are
produced by inputting $(S_{+},S_{-})$=(0,1) or (-1,0), whereas excited SVs are
produced by inputting $(S_{+},S_{-})$=$(n,n+1)$, where $n$ is an integer and
satisfies $n\neq-1$ or 0. As for the fundamental MMs and their excited states,
the ansatz is defined as
\begin{equation}
\phi_{\pm}^{(0)}=A_{1}r^{|S_{1}|}\exp(-\alpha_{1}r^{2}\pm iS_{1}\theta)\mp
A_{2}r^{|S_{2}|}\exp(-\alpha_{2}r^{2}\mp iS_{2}\theta),\label{ESMM}%
\end{equation}
where $A_{1,2}$ and $\alpha_{1,2}$ are arbitrary real constants. $S_{1,2}$ are
the topological charge numbers and satisfy $S_{2}=S_{1}+1$. Fundamental MMs
can be created when $S_{1}=-1$ or 0, while excited MMs are created when
$S_{1}$ is assigned other integer values. Because the interactions, which
includes the self-interaction and the cross-interaction, in the spinor BECs
are tuned to be a homogeneous repulsive interaction. As expected, stable
excited SVs and MMs are found in this system. Typical examples of excited
states of SVs and MMs corresponding to $S_{+}=5$ and $S_{1}=5$ are displayed
in Fig. \ref{ESVEMM}. The characteristic of these excited states are in
accordance with their counterparts found in Ref.
\cite{Chunqing2018,Rongxuan2018}. However, the excited states found in
previous systems, which were built via the modulation of local and nonlocal
repulsion, are only stable up to $S_{1}$ or $S_{+}$ up to 5. In contrast, in
the current system, excited modes can be found with $S_{+}$ or $S_{1}>5$ for
appropriate values of the control parameters of $(D,g,\gamma)$. In the case of
self-attractive BECs with confined SO coupling, the soliton was formed with
the help not only from the confined SO coupling but also from the
self-attraction \cite{Yongyao20192}. Under this circumstance, the confined SO
coupling dose not feature a purely effect. However, in the current system, if
we replace the confined SO coupling with the homogeneous SO coupling, the BECs
diffuses by the repulsion. Therefore, the localized modes found in the current
system purely depend on the trapping ability of the confined SO coupling. To
study the dependence of the trapping ability of the confined SO coupling on
the control parameters of the system, we plot the size, the chemical potential
and the energy of the nonlinear mode, i.e., $(R,\mu,E)$, as functions of
$(D,g,\gamma)$ in Fig. \ref{char3}. Here, the size of the nonlinear mode is
defined as
\begin{equation}
R={\left(  {\frac{{\int\mathbf{r}{{^{2}}n\left(  \mathbf{r}\right)  }%
}d\mathbf{r}}{{\int{n\left(  \mathbf{r}\right)  }}d\mathbf{r}}}\right)
^{\frac{1}{2}},}\label{R}%
\end{equation}
where $n\left(  \mathbf{r}\right)  =|\phi_{+}(\mathbf{r})|^{2}+|\phi
_{-}(\mathbf{r})|^{2}$ is the total density distribution of the nonlinear
mode. For convenience, we use the fundamental mode to characterize this dependence.

Figs. \ref{char3} (a1, a2, a3) display the size $R$, the chemical potential
$\mu$, and the total energy $E$ of fundamental SVs with $\gamma=1$,
respectively, as functions of the SO coupling confinement size, $D$, for
different strengths of the repulsive interaction $g$. In Fig. \ref{char3}(a1),
the size of the nonlinear mode, which is denoted by $R\left(  D\right)  $,
totally increases with increasing $D$ when $D>10$. This result indicates that
the size of the nonlinear mode expands due to the repulsion if the trapping
from the confined SO coupling becomes weaker. When $D\rightarrow\infty$, the
confinement disappears, the SO coupling becomes homogeneous, and the nonlinear
mode may decay to an infinite-size linear mode in 2D free space with a
homogeneous SO coupling background. However, if the confined area $D$ is too
small, then it cannot provide enough area for trapping the nonlinear mode. In
this circumstance, the size of the nonlinear mode starts to expand again. When
$D\rightarrow0$, the SO coupling disappears from the 2D space, which results
in decay of the nonlinear mode to a linear mode in a purely 2D free space.
Between these two limits ($D\rightarrow0$ and $D\rightarrow\infty$), an
optimized value of $D$ appears at $D\approx10$, at which the nonlinear modes
are trapped with the smallest size. The existence of such an optimized value
of $D$ is also predicted in Ref. \cite{Yongyao2019}. In Figs. \ref{char3}(a2,
a3), $\mu,E\rightarrow0$ and $-0.5$ when $D\rightarrow0$ and $\infty$, which
are in accordance with the values of the chemical potential and energy of the
linear mode in 2D free space with and without a homogeneous SO coupling
background \cite{Sakaguchi2014}.

Figs. \ref{char3}(b1, b2, b3) display the $(R,\mu,E)$ of fundamental SVs as
functions of $g$ with $\gamma=1$ and different values of $D$. The dependence
between these characters and $g$ can be naturally understood. Fig.
\ref{char3}(b1) indicates that an increase in the strength of the repulsion of
BECs may increase the size of the nonlinear mode. Figs. \ref{char3}(b2, b3)
imply that the functions of $\mu(g)$ and $E(g)$ satisfy the anti-VK
(Vakhitov-Kolokolov) criterion, which is a necessary stability condition for
nonlinear modes with a repulsive interaction \cite{antiVK}.

Finally, Fig. \ref{char3} (c1, c2, c3) present the $(R,\mu,E)$ of the two
types of nonlinear modes, fundamental SVs and MMs, as functions of $\gamma$
for fixed values of $D$ and $g$. The figures show that these functions all
increase with increasing $\gamma$. $(R,\mu,E)_{\mathrm{MM}}$ is smaller or
greater than $(R,\mu,E)_{\mathrm{SV}}$ when $\gamma<1$ or $\gamma>1$,
respectively. Hence, intersections are found between the functions for SVs and
MMs at $\gamma=1$. This finding shows that SVs and MMs are degenerate at this
point, which is the same finding as that in previous works
\cite{Sakaguchi2014,Rongxuan2018}: the SVs and MMs are degenerate under the
condition of a Manakov type \cite{VSS}. This result also explains why we can
use SVs with $\gamma=1$ to identify the characteristics of the confined SO
coupling in Figs. \ref{char3} (a1, a2, a3) and (b1, b2, b3).

\begin{figure}[tbh]
{\includegraphics[width=18cm]{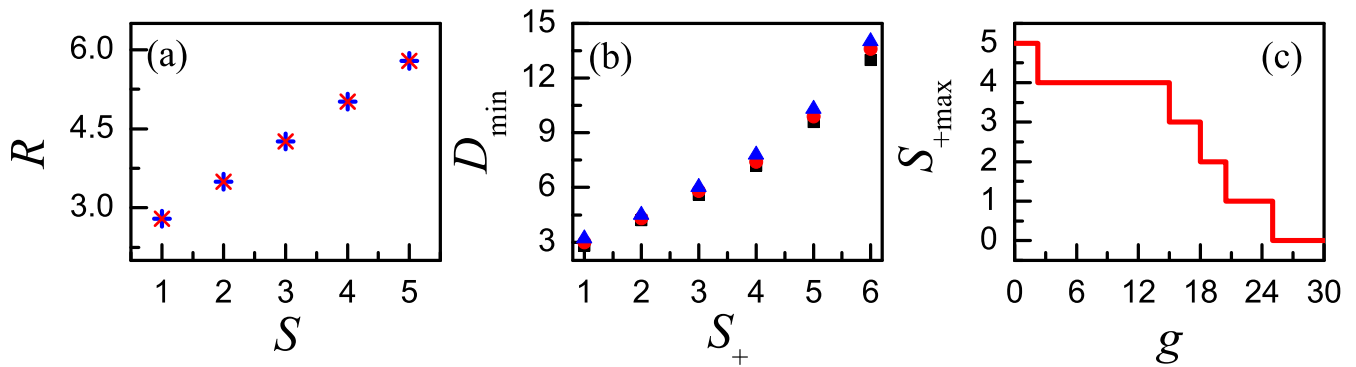}}\caption{(color online) (a) The size
of the total density pattern of the excited SVs (marked with \textquotedblleft%
$\times$\textquotedblright symbols) and MMs (marked with \textquotedblleft%
$+$\textquotedblright symbols) with different topological number $S$; here,
$\left(  N,g,D\right)  =\left(  1,2,10\right)  $. (b) The dependence of the
minimum SO coupling confinement size, $D_{\min}$, on the topological number
$S_{+}$ for different repulsive strengths $g$ ($g=0$ for solid black squares,
$g=2$ for solid red circles and $g=4$ for solid blue triangles).\ (c) The
trapping ability for the maximum topological number $S_{+\max}$ versus the
degree of the nonlinear strength $g$; here, $\left(  N,\gamma,D\right)
=\left(  1,1,10\right)  $. }%
\label{Esconfined}%
\end{figure}\ \ 

The above discussion on the characteristics of the confined SO coupling was
based on the fundamental mode. Because this system supports the existence of
the excited mode, a nontrivial discussion related to the capacity of the
confined SO coupling for the excited modes needs to be addressed by the
current paper. Because the excited SVs and MMs are also degenerate at
$\gamma=1$, which is the same as their fundamental counterparts, we will adopt
excited SVs with $\gamma=1$ in the following discussion for convenience. The
excited modes in this system are characterized by the topological number
(i.e., $S_{+}$ for excited SVs or $S_{1}$ for excited MMs). Fig.
\ref{Esconfined}(a) shows that the size of the total density pattern of the
excited SVs and MMs increases with the number of positive $S_{+}$ and $S_{1}$.
The complete overlap of $R(S_{+})$ and $R(S_{1})$ demonstrates the validity of
using excited SVs to represent excited MMs at $\gamma=1$ to study the capacity
of the confined SO coupling for the excited modes.

Numerical studies find that there is a minimum area, i.e., a threshold of
confined area, for the SO coupling confinement, namely, $D_{\min}$, for
excited modes with different values of $S_{+}$. Fig. \ref{Esconfined}(b) shows
the dependence of $D_{\min}$ on the charge number $S_{+}$ for different values
of $g$. When $D>D_{\min}(S_{+})$, a stable excited state with charge number
$S_{+}$ can be found in the system. When $D<D_{\min}(S_{+})$, an excited mode
with charge number $S_{+}$ cannot exist in the system. The numerical
simulation shows that all the inputs with charge number greater than $S_{+}$
converge to the fundamental mode. Fig. \ref{Esconfined}(b) indicates that
$D_{\min}(S_{+})$ increases as $S_{+}$ and $g$ increase, which can be
naturally understood because higher excited modes with larger values of $g$
require larger thresholds for the confined area because they possess larger
values of $R$. On the other hand, Fig. \ref{Esconfined}(b) can also be applied
to estimate the number of excited modes trapped by the confined SO coupling.
For example, if we select $g=2$ and $D=10$, which satisfies $D>D_{\min}%
(S_{+}=5)$, such a confined area can contain the excited mode with topological
number up to $S_{+}=5$. According to this application, for a fixed value of
$D$, there is a maximum topological number $S_{+\max}$, which denotes the
number of excited modes existing in this setting. Hence, $S_{+\max}$ can be
use to identify the capacity of confined SO coupling with a fixed value of
$D$. Fig. \ref{Esconfined}(c) manifests the dependence of the maximum
topological number $S_{+\max}$ and the repulsive strength $g$ for a fixed
value of $D$. As discussed above, since an increase in $g$ can expand the size
of the excited nonlinear mode, one can see that the capacity of the confined
SO coupling decreases with increasing $g$.

\section{Nonlinear mode in confined SO coupling with a moving reference frame}

Unlike the usual system without SO coupling, a system with SO coupling does
not obey Galilean invariance. Hence, studying the mobility of the nonlinear
mode is another nontrivial issue for a system with SO coupling. Reference
\cite{Sakaguchi2014} reported that only one type of stable nonlinear mode was
found when the system only moves in the $y$ direction up to a threshold
velocity. For the current system, we will demonstrate that more than one type
of stable nonlinear mode can be found; additionally, a nonlinear mode can be
found to be stable when the system moves in the $x$ direction.

Here, we assume stable transfer of the nonlinear mode by the moving SO
coupling profile, which corresponds to
\begin{align}
&  \left(  x^{\prime},y^{\prime}\right)  \rightarrow\left(  x-v_{x}%
t,y-v_{y}t\right)  ,\nonumber\\
&  \lambda^{\prime}(r^{\prime})\rightarrow\lambda(r),\nonumber\\
&  \tilde{\Psi}_{\pm}(\mathbf{r^{\prime}},t)=\Psi_{\pm}(\mathbf{r}%
,t),\nonumber\\
&  \Psi_{\pm}^{\prime}(\mathbf{r}^{\prime},t)=\tilde{\Psi}_{\pm}%
(\mathbf{r}^{\prime},t)\exp\left[  {\frac{i}{2}}\left(  v_{x}x^{\prime2}%
+v_{y}y^{\prime2}\right)  \right]  .
\end{align}
Eqs. (\ref{Model}) are transformed into the following forms:%
\begin{align}
i\frac{\partial{\Psi}_{+}^{\prime}}{\partial t}  &  =-\frac{1}{2}%
\nabla^{\prime2}{\Psi}_{+}^{\prime}+g\left(  \left\vert \Psi_{+}^{\prime
}\right\vert ^{2}+\gamma\left\vert \Psi_{-}^{\prime}\right\vert ^{2}\right)
\Psi_{+}^{\prime}+\lambda^{\prime}\left(  \frac{\partial}{\partial x^{\prime}%
}-i\frac{\partial}{\partial y^{\prime}}\right)  \Psi_{-}^{\prime}%
+\frac{e^{-i\theta^{\prime}}}{2}{\frac{d\lambda^{\prime}}{dr^{\prime}}}%
\Psi_{-}^{\prime}+\lambda^{\prime}(r^{\prime})\left(  iv_{x}+v_{y}\right)
\Psi_{-}^{\prime},\nonumber\\
i\frac{\partial{\Psi}_{-}^{\prime}}{\partial t}  &  =-\frac{1}{2}%
\nabla^{\prime2}{\Psi}_{-}^{\prime}+g\left(  \left\vert \Psi_{-}^{\prime
}\right\vert ^{2}+\gamma\left\vert \Psi_{+}^{\prime}\right\vert ^{2}\right)
\Psi_{-}^{\prime}-\lambda^{\prime}\left(  \frac{\partial}{\partial x^{\prime}%
}+i\frac{\partial}{\partial y^{\prime}}\right)  \Psi_{+}^{\prime}%
-\frac{e^{i\theta^{\prime}}}{2}{\frac{d\lambda^{\prime}}{dr^{\prime}}}\Psi
_{+}^{\prime}-\lambda^{\prime}(r^{\prime})\left(  iv_{x}-v_{y}\right)
\Psi_{+}^{\prime}. \label{model_moving}%
\end{align}
Obviously, the last term in Eqs. (\ref{model_moving}), which is modulated by
$\lambda^{\prime}(r^{\prime})$, makes the nonlinear modes in the current
system differ from their counterparts in the system with homogeneous SO coupling.

\begin{figure}[ptb]
{\includegraphics[width=1.0\columnwidth]{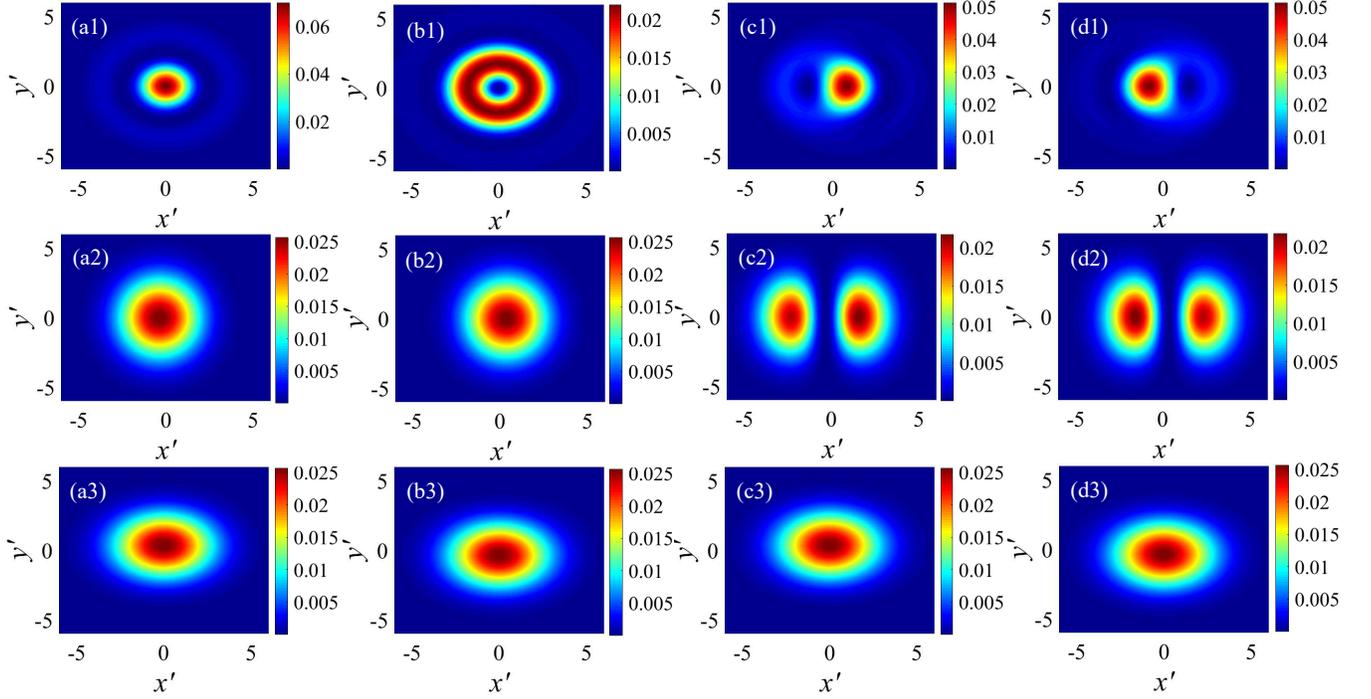}}\caption{(Color online)
Density distribution of the nonlinear mode in the moving reference frame.
Figs. (a1, b1, c1, d1) shows the system with $v_{x}=v_{y}=0$, and Figs. (a2,
b2, c2, d2) shows the system moving in the $y$-direction with $v_{x}=0$ and
$v_{y}=0.5$. Figs. (a3, b3, c3, d3) shows the system moving in the
$x$-direction with $v_{x}=0.5$ and $v_{y}=0$. The first through fourth columns
present the density distribution for each component obtained by inputting an
SV guess and an MM guess into Eqs. (\ref{ESSV}) and (\ref{ESMM}),
respectively. Here, we fixed $\left(  D,\gamma,g\right)  =\left(
10,1,2\right)  $.}%
\label{movingF}%
\end{figure}

Here, a stationary solution in the moving reference frame is obtained by
solving Eqs. (\ref{model_moving}) by means of the ITM. The stabilities of the
solution are checked via direct simulation using Eqs. (\ref{model_moving}).
The inputs still adopt the definitions in Eqs. (\ref{ESSV}) and (\ref{ESMM})
for the two types of nonlinear modes. For convenience, we will fix $D=10$ and
$\gamma=1$ in this section. Typical examples of the stable nonlinear modes,
which are generated by inputting Eqs. (\ref{ESSV}) and (\ref{ESMM}) with
$S_{+}=S_{1}=0$, for states with different values of $v_{x}$ and $v_{y}$ are
displayed in Fig. \ref{movingF}.

In the limit of $v_{x}=v_{y}=0$, which is the same as the system in the
quiescent reference frame, as expected, stable fundamental SV modes and MMs
are generated by inputting different types of guesses. However, once $v_{x}$
or $v_{y}$ differ from $0$, only nonlinear modes with
\[
N_{+}=\int|\Psi_{+}^{\prime}(\mathbf{r^{\prime}})|^{2}d\mathbf{r^{\prime}%
}=\int|\Psi_{-}^{\prime}(\mathbf{r^{\prime}})|^{2}d\mathbf{r^{\prime}}=N_{-}%
\]
are found in the system. In the case of $v_{y}\neq0$ and $v_{x}=0$, two types
of stationary modes are found by inputting two types of guesses. The SV guess
in Eq. (\ref{ESSV}) produces a single hump profile density distribution for
each component. The density distributions of the two components deviating from
the axis of $x^{\prime}=0$ feature a mirror symmetry about this axis, which is
similar to the findings of nonlinear modes in the mobility system with
$v_{y}\neq0$ in Refs. \cite{Sakaguchi2014} and \cite{Yongyao2017}. In
contrast, the MM guess in Eq. (\ref{ESMM}) generates a stable nonlinear mode
different from the counterpart produced by the SV guess. The density
distributions of such nonlinear modes have a double-hump structure. The center
between the two humps is exactly located at the axis of $x^{\prime}=0$.
Typical examples of these two types of nonlinear modes are displayed in the
second row of Fig. \ref{movingF}. Finally, in the case of $v_{x}\neq0$ and
$v_{y}=0$, only one type of nonlinear mode can be produced regardless of
whether an SV guess or an MM guess are input. The density distribution of this
nonlinear mode has a single hump structure. It is similar to the nonlinear
mode produced by the SV guess in the case of $v_{x}=0$ and $v_{y}\neq0$,
whereas its mirror symmetry axis changes to $y^{\prime}=0$. Typical examples
of this type of nonlinear mode are shown in the third row of Fig.
\ref{movingF}.

The system with homogeneous SO coupling (refer to Ref. \cite{Sakaguchi2014}),
which presented the stable MM moves only in the $y$ direction up to a
threshold velocity and the SV mode is hardly moving. In our system, we find
different types of stationary mobility modes when the moving velocity is along
the $x$ and $y$ directions. In the case of moving along the $y$ direction, two
types of nonlinear modes can be found by inputting different types of guesses,
while in the case of moving along the $x$ direction, only one type of
nonlinear mode can be found regardless of what type of guess is input. This
result reveals that the anisotropic moving characteristics for the confined 2D
SO coupling, which are different from the counterpart produced for the
homogeneous SO coupling.

\section{Conclusion}

The objective of this work is to construct several types of self-trapping
nonlinear modes in the 2D model of binary Bose-Einstein condensates (BECs)
with a repulsive interaction and spatially confined spin-orbit (SO) coupling.
Spatially confined SO coupling can be induced in spinor BECs by illuminating
them with a Gaussian laser beam. Such a confined effect can make the SO
coupling feature a trapping capacity for the nonlinear mode. The most
essential finding is that compared to the attractive interaction, excited
semi-vortex (SV) modes and mixed modes (MMs) with higher topological charge
numbers can be stabilized by the confined SO coupling in this setting. The
trapping ability versus the degree of the repulsive strength as well as the
topological charge number are systematically identified in the paper. The
minimum SO coupling confined size, $D_{min}$, is found for excited modes with
different topological charge numbers and different strengths of repulsion. A
nontrivial discussion on the capacity of the SO coupling with fixed confined
size, which defines how many excited modes can be contained in this confined
area, is presented in the paper. Another nontrivial discussion on the
stability of the nonlinear modes in a moving system is also considered. We
find that different types of stationary mobility modes can be stabilized when
the SO coupling moves in the $x$- and $y$-directions. This finding indicates
that the system with moving confined SO coupling features a different
anisotropic character compared to its counterpart with moving homogeneous SO coupling.

\section{Acknowledgments}

This work was supported by NNSFC (China) through grant Nos. 11905032,
11874112, and 11575063, the Foundation for Distinguished Young Talents in
Higher Education of Guangdong through grant Nos. 2018KQNCX279 and
2018KQNCX009, and the Special Funds for the Cultivation of Guangdong College
Students Scientific and Technological Innovation, No. pdjh2019b0514.

\textbf{Compliance with ethical standards}

\textbf{Conflict of interest} The authors have declared that no conflict of
interest exists.

\textbf{Ethical standards} This Research does not involve Human Participants
and/or Animals.

\end{document}